\documentclass[10pt]{aastex}
\usepackage{hyperref}

\begin{document}
\title{DAMEWARE: A web cyberinfrastructure for astrophysical data mining}

\author{Massimo Brescia, Stefano Cavuoti}
\affil{Astronomical Observatory of Capodimonte, INAF, Napoli, Italy}
\author{Giuseppe Longo\altaffilmark{1}}
\affil{Department of Physics, University Federico II, Napoli, Italy}
\altaffiltext{1}{Astronomical Observatory of Capodimonte, INAF, Napoli, Italy}
\author{Alfonso Nocella, Mauro Garofalo, Francesco Manna, Francesco Esposito, Giovanni Albano, Marisa Guglielmo, Giovanni D'Angelo, Alessandro Di Guido}
\affil{Department of Computer Science and Engineering, University Federico II, Napoli, Italy}
\author{George S. Djorgovski, Ciro Donalek, Ashish A. Mahabal}
\affil{Department of Astronomy, California Institute of Technology, Pasadena, USA}
\author{Matthew J. Graham}
\affil{Center for Advanced Research in Computing, California Institute of Technology, Pasadena, USA}
\author{Michelangelo Fiore}
\affil{LAAS - CNRS Toulouse, France}
\and
\author{Raffaele D'Abrusco}
\affil{Harvard Smithsonian Center for Astrophysics, Harvard, USA}


\graphicspath{{./}}

\begin{abstract}
Astronomy is undergoing through a methodological revolution triggered by an unprecedented wealth of
complex and accurate data. The new panchromatic, synoptic sky surveys require advanced tools for
discovering patterns and trends hidden behind  data which are both complex and of high dimensionality.
We present DAMEWARE (DAta Mining \& Exploration Web Application \& REsource): a general purpose,
web-based, distributed data mining environment developed for the exploration of large data sets,
and finely tuned for astronomical applications.
By means of graphical user interfaces, it allows the user to perform classification, regression or
clustering tasks with machine learning methods.
Salient features of DAMEWARE include its capability to work on large datasets with minimal human
intervention, and to deal with a wide variety of real problems such as the classification of globular clusters in the
galaxy NGC1399, the evaluation of photometric redshifts and, finally, the identification of candidate Active Galactic Nuclei
in multiband photometric surveys. In all these applications, DAMEWARE allowed to achieve better results than
those attained with more traditional methods.
With the aim of providing potential users with all needed information, in this paper we briefly
describe the technological background of DAMEWARE, give a short introduction to some relevant aspects of data mining, followed by a
summary of some science cases and, finally, we provide a detailed description of a template use case.
\end{abstract}

\section{Introduction}

Astronomy has recently become a data rich science and not only data volumes and data rates
are growing exponentially, closely following Moore's law \citep{szalay2006} but, at the same time,
there are also significant increases in data complexity and data quality.
For instance, the data provided by the new panchromatic, synoptic surveys often consists of tables
containing many hundreds of parameters and quality flags for billions of objects.
These parameters are often highly correlated and carry redundant information which introduce hard to disentangle ambiguities.
In addition, most times, these tables are plagued by a large fraction of missing (not a number or NaN)
data which need to be taken into account properly.

It is not just this new data abundance that is fueling
this revolution, but also the Internet-enabled data access and the extensive data re-use.
The informational content of the modern data sets is in fact so high as to render archival research
and data mining mandatory since, in most cases, the
researchers who took the data tackle just a small fraction of the science that is enabled by it.
The main implication being that the same data sets (or specific subsets of them) are often used by many different teams
to tackle different problems which were not among the main goals of the original surveys.
This multiplication of experiments requires optimal strategies for data extraction and transfer (from the
data repositories to the final user) and for data processing and mining.

A first response of the astronomical community to these challenges was the Virtual Observatory (VO) which was initially
envisioned as a complete, distributed (Web-based) research environment for astronomy with large and complex data sets,
to be implemented by federating geographically distributed data and
computing facilities, as well as the necessary tools and related expertise \citep{brunner2001,george2002}.
The VO is nowadays a world-wide enterprise, with a number of national and international VO
organizations federated under the umbrella of the International Virtual Observatory Alliance
(IVOA\footnote{ \url{http://ivoa.net}}).
The VO implementation, however, has so far focused mainly on the production of the necessary data
infrastructure, interoperability, standards, protocols, middleware, data discovery services, and
has produced only a limited subset of data exploration and analysis services.
Very little has so far been done in the production of tools capable to explore large data sets
and to extract in a semiautomatic way, patterns and useful information from a wealth of data
which goes well beyond the capabilities of individual analysis.
This process is usually called either Knowledge Discovery in Databases or \textit{Data Mining} with very
small if any, semantic differences between the two wordings.

In spite of the fact that in the last few years \textit{Data Mining} (DM) seems to have become immensely
popular in the astronomical community (which has begun to label as data mining any sort of
data querying, data analysis or data visualization procedure), true \textit{Data Mining} -i.e. the extraction of
useful information from the data with automatic methods- is still quite uncommon.
This is probably due to the fact that DM is a complex and non deterministic process where the optimal results
can be found only on a trial-and-error base, by comparing the output of different methods and
of different experiments performed with the same method.
This implies that, for a specific problem, DM requires a lengthy fine tuning phase which
is often not easily justifiable to the eyes of a non-experienced user.
Furthermore, in order to be effective, DM requires a good understanding of the mathematics
underlying the methods, of the computing infrastructures, and of the complex workflows which
need to be implemented.
Most casual users (even in the scientific community) are usually not willing to make the effort
to understand the process, and prefer recourse to traditional and less demanding approaches
which are far less powerful but often much more user friendly \citep{hey2009}.
This however, will become more and more difficult in the future when DM will become an unavoidable necessity.

Many  Data Mining packages are nowadays available to the scientific community,
from desktop applications (i.e. packages to be downloaded and installed on user
local machine)
to web-based tools (services and applications which can be remotely executed from a simple browser).
To the first group, to quote just a few, belong Knime\footnote{\url{http://www.knime.org/}},
Orange\footnote{\url{http://orange.biolab.si}}, Weka\footnote{\url{http://www.cs.waikato.ac.nz/~ml/weka/}}
and RapidMiner\footnote{\url{http://rapid-i.com/content/view/181/196/}}.
While VOStat\footnote{\url{http://astrostatistics.psu.edu/vostat/}}
and the DAMEWARE\footnote{\url{http://dame.dsf.unina.it/dameware.html}}, described here,
belong to the second group.
Many of these tools have been the subject of a review study carried out to determine which of the wide
variety of available data mining, statistical analysis and visualization applications and algorithms could
be most effectively used by the astrophysical community \citep{donalek2011}.
The main result of this study being that most of these tools fail to scale when applied even to moderately large (hundreds of thousands records) data sets.

DAMEWARE (DAta Mining \& Exploration Web Application REsource) was conceived and engineered in 2007
to enable a generic user to perform data mining and exploratory experiments on large Data
Sets (of the order of few tens of GBytes) and, by exploiting web $2.0$ technologies, it offers several tools which can be seen
as working environments where to choose data analysis functionalities such as clustering,
classification, regression, feature extraction etc., together with models and algorithms.
As it will be shown in what follows, under DAMEWARE, any user can setup, configure and
execute experiments on his own data, on top of a virtualized computing infrastructure, without the
need to install any software on his local machines.

DAMEWARE has been offered to the public since early $2012$.
During its commissioning period, ended in August $2013$, about $100$ scientists from $27$ different countries
registered as users and performed many different experiments.
In the same period, the project web site hosted $\sim12,000$ independent accesses.
Furthermore, the various functionalities and models offered by DAMEWARE have been extensively tested on
real scientific cases and the results are discussed in $25$ publications, among which are $10$ refereed papers,
$3$ contributes to volumes and $12$ proceedings of international workshops.

This paper merges two aspects: first of all, it intends to provide a concise
technical description of DAMEWARE; second it provides the interested reader with
a quick overview of the functionalities and with a worked out template use case.
In the next section (Sect.~\ref{sect:dame:design}) we describe the design and the architecture of
DAMEWARE web application. Sect.~\ref{sect:DM} gives the methodological background behind DAMEWARE
while in Sect.~\ref{sect:usingdameware} we describe in some detail how DAMEWARE works, describing the data
preprocessing (Sect.~\ref{sect:data}), the experiments and the post reduction (\ref{sect:visualization}).
In Sect. \ref{sect:science} we shortly outline some previous applications of DAMEWARE which can be used by the interested reader to better understand the potentialities of  DAMEWARE and,
in order to better exemplify the workflow involved in a typical DAMEWARE experiment, in Sect.~\ref{sect:app} we present, as template use case, the evaluation of photometric redshifts for a sample of galaxies
used by our team for the PHAT1 contest \citep{hildebrandt2010,cavuoti2012}.
Finally in Sect.~\ref{sect:conclusion} we outline some key points of the discussion and draw some conclusions.

Readers who are not interested in the technical aspects and/or who have enough experience in data mining can skip the first sections and move directly to Sect.~\ref{sect:science} and \ref{sect:app}.

\section{DAMEWARE design and architecture}\label{sect:dame:design}
From a technical point of view, DAMEWARE is what is called a Rich Internet Application (RIA, \citealt{bozzon2010}),
consisting of web applications having the traditional interaction and interface features of computer programs,
but usable via  web browsers.
The main advantage of using web applications is that the user is not required to install program clients on his
desktop, and has the possibility to collect, retrieve, visualize and organize the data, as well as configure and
execute the data mining applications through his web browser.
An added value of such approach is the fact that the user does not need to directly
access large computing and storage  facilities, and can transparently perform his experiments
exploiting computing networks and archives located worldwide, requiring only a local laptop
(or even a smartphone or a tablet) and a network connection.

Most available web based data mining services run synchronously and this implies that they
execute jobs during a single HTTP transaction. In other words, all the entities in the chain
(client, workflow engine, broker, processing services)  must remain up for the whole duration
of the activity; if any component stops, the context of the activity is lost.
Obviously, this approach does not match the needs of long-run tasks which are
instead the rule when dealing with large data sets.
For this reason, DAMEWARE offers asynchronous access to the infrastructure tools,
thus allowing the running of activity jobs and processes outside the scope of any particular
web service operation and without depending on the user connection status.
In other words, the user, via a simple web browser, can access the application resources and
has the possibility to keep track of his jobs by recovering related information (partial/complete results)
at any moment without being forced to maintain open the communication socket.

From the software development point of view, the baselines behind the engineering design of DAMEWARE were:
\begin{itemize}
\item	\emph{Modularity}: software components with standard interfacing, easy to be replaced;
\item	\emph{Standardization}: in terms of information I/O between user and infrastructure, as well as
between software components (in this case based on the XML-schema);
\item	\emph{Interoperability of data}: obtained by matching VO requirements (standards and formats);
\item	\emph{Expandability}: many parts of the framework will be required to be enlarged and updated
along their lifetime.
This is particularly true for the computing architecture, framework capabilities, GUI (Graphical User Interface)
features, data mining functionalities and models (this also includes the integration within the framework of end
user proprietary algorithms);
\item	\emph{Asynchronous interaction}: the end user and the client-server mechanisms do not require a synchronous interaction.
By using the Ajax (Asynchronous Javascript and XML, described in \citealt{garrett2005}) mechanism, the web applications can
retrieve data from the server running asynchronously in  background, without interfering with the display and behavior of the
existing page;
\item	\emph{Language-independent Programming}. This basically concerns the APIs (Application Programming Interface) forming
the data mining model libraries and packages. Although most of the available models and algorithms were internally implemented,
this is not considered as mandatory, since it is possible to re-use existing tools and libraries, to integrate end user tools, etc.
To this aim, the Suite includes a Java based standard wrapping system to achieve the standard interface with multi-language APIs;
\item \emph{Hardware virtualization}. DAMEWARE is independent from the hardware deployment platform (single or multi processor, grid etc.).
\end{itemize}

To complete this short summary, we wish to note that DAMEWARE also offers  a downloadable Java desktop
application (called DMPlugin\footnote{\url{http://dame.dsf.unina.it/dameware.html\#plugin}}), which allows the end
users to extend the original library of available data analysis tools by plugging-in, exposing to the community
and executing their own code in a simple way, just by uploading into the framework their programs, without any
restriction on the native programming language.

Furthermore the many-core -the new parallel processing paradigm recently replacing the multi-core concept-  hardware platform hosting
the web application, supports the possibility of running parallel
programs \citep{barsdell2010}, via a dedicated Graphical Processing Unit (GPU, \citealt{nvidia2012}) $K20$ device.

Finally, we wish to stress that, in order to be as user friendly as possible, special care was given to
the documentation, both technical and user oriented, which is accessible through the website.

\section{Data Mining \& DAMEWARE}\label{sect:DM}

In order to better understand the problems related with data mining in general and with astronomical data mining in particular,
it is necessary to spend a few words on the distinction between the well known concept of observed astronomical space and the
so called measurements astronomical parameter space as defined in \citep{djorgovski2012}.

\subsection{Preparing the data}
Every astronomical observation, surveys included, covers some finite portion of the Observable Parameter Space (OPS), whose
axes correspond to the observable quantities, e.g., flux wavelength, sky coverage, etc. (see below).
Every astronomical observation or a set thereof, surveys included, subtends a multi-dimensional volume (hypervolume) in this
multi-dimensional parameter space.
The dimensionality of the OPS is given by the number of characteristics that can be defined for a given type of observation.
Along some axes, the coverage may be intrinsically discrete, rather than continuous.
An observation can be just a single point along some axis, but have a finite extent in others.
A correct characterization of the OPS is useful for many applications but for the purposes of Data Mining, it is
necessary to introduce the  quite different concept of Measurement Parameter Space (MPS).
Catalogs of sources and their measured properties can be represented as points (or vectors) in the MPS.
Every measured quantity for the individual sources has a corresponding axis in the MPS.
But differently from OPS, some can be derived from the primary measured quantities; for example, if
the data are obtained in multiple bandpasses, we can form axes of flux ratios or colors; a difference of
magnitudes in different apertures forms a concentration index; surface brightness profiles of individual
objects can be constructed and parametrized, e.g., with the Sersic index; and so on.
Some parameters may not even be representable as numbers, but rather as labels; for example, morphological
types of galaxies, or a star vs. a galaxy classification.
While OPS represents the scope and the limitations of observations, MPS is populated by the detected sources
and their measured properties. It describes completely the content of catalogs derived from the surveys.
Each detected source is then fully represented as a feature vector in the MPS (\emph{features} is a commonly
used computer-science term for what we call measured parameters here).
Modern imaging surveys may measure hundreds of parameters for each object, with a corresponding (very high)
dimensionality of the MPS which is a curse for any data mining applications and not only because algorithms
scale badly with an increasing number of features, but also because for increasing number of dimensions the
average density of information decreases (i.e. the n-dimensional table becomes more and more sparsely populated)
thus increasing the level of noise.
In other words, to the contrary of  what is normally perceived, an high number of features is very often
a nuisance rather than a help \citep{djorgovski2012}.

Missing data or NaN, change the dimensionality of the affected records in the MPS, a fact which is not easy to deal
with any DM method. Astronomical data present a further level of complexity since the missing information can be of
two types.
In the simplest case, the missing data is truly a NaN: for instance objects which have not been observed in one photometric
band and therefore correspond to truly missing information.
In the second case, the missing data can be, for instance, an object which has been observed but not detected in a
given band.
The non detection conveys some information on the physical properties of the object itself and cannot be plainly
ignored.
In this case, it might be convenient to substitute the missing data with some relevant information such as, in the case
of the previous example, an upper limit to the flux.

It may also happen that the information content in a single table is not homogeneous, i. e. attributes may be of different
types, such as numerical variables mixed with categorical ones.
This level of diversity in the internal information can also be related to different format type of data sets,
such as tables registered in ASCII code \citep{ansi1977}, CSV (Comma Separated Values, \citealt{repici2010}) or
FITS (text header followed by binary code of an image, \citealt{wells1981}).
In order to reach an efficient and homogeneous representation of the data set to be submitted to ML systems, it is mandatory
to preliminarily take care of the data format, in order to make them intelligible by the processing framework.
In other words, it is crucial to transform features and force them into a uniform representation before starting the
DM process.
In this respect real working cases are, almost always, quite different among themselves.
Let us, for instance, think of time series  (coming from any sensor monitoring acquisition) where data are collected in
a single long sequence, not simply divisible, or of raw data that could be affected by noise or  aberration factors.

Furthermore, most practical DM methods based on the ML paradigm, when dealing with massive data sets make an intensive use
of so-called \textit{meta-data},\footnote{A \textit{meta-datum} - from the Greek \textit{meta}, or \textit{over, after}
and the Latin \textit{datum} or \textit{information}. -
is the information that describes a whole set of data (Guenther et al. 2004).}
another category of data representation, based on partial ordering or equivalently generalization/specialization relations.
Fields of a meta-data collection are composed of information describing resources and quick notes related to the referred
original data, able to improve their fast visibility and access.
They also provide the primary information retrieval, indexed through description fields, usually formatted as records
(pattern of labels).

Feature selection \citep{guyon2003} consists of identifying which among the various input parameters (also called
\textit{features} in DM terminology) carry the largest amount of information.
This allows one to exclude from subsequent analysis the less significant features with the twofold purpose of decreasing
the dimensionality of the problem (thus reducing the noise in the data), and of improving the performance in terms of computing
time.
Feature selection requires a large number of independent experiments and a lengthy comparison procedure. In practice specific
combinations of input features are created and submitted as independent runs of the method and only those combinations which
do not cause a significant loss in performance are maintained.

Many different approaches are possible and we shall not enter into any detail but the interested reader can refer to:
\citep{md1,md2,vashist2012} for the missing data problem and to \citep{fs1}  for an in depth analysis of the possible
feature selection procedures.
\begin{figure*}
\centering
  \includegraphics[width=14cm]{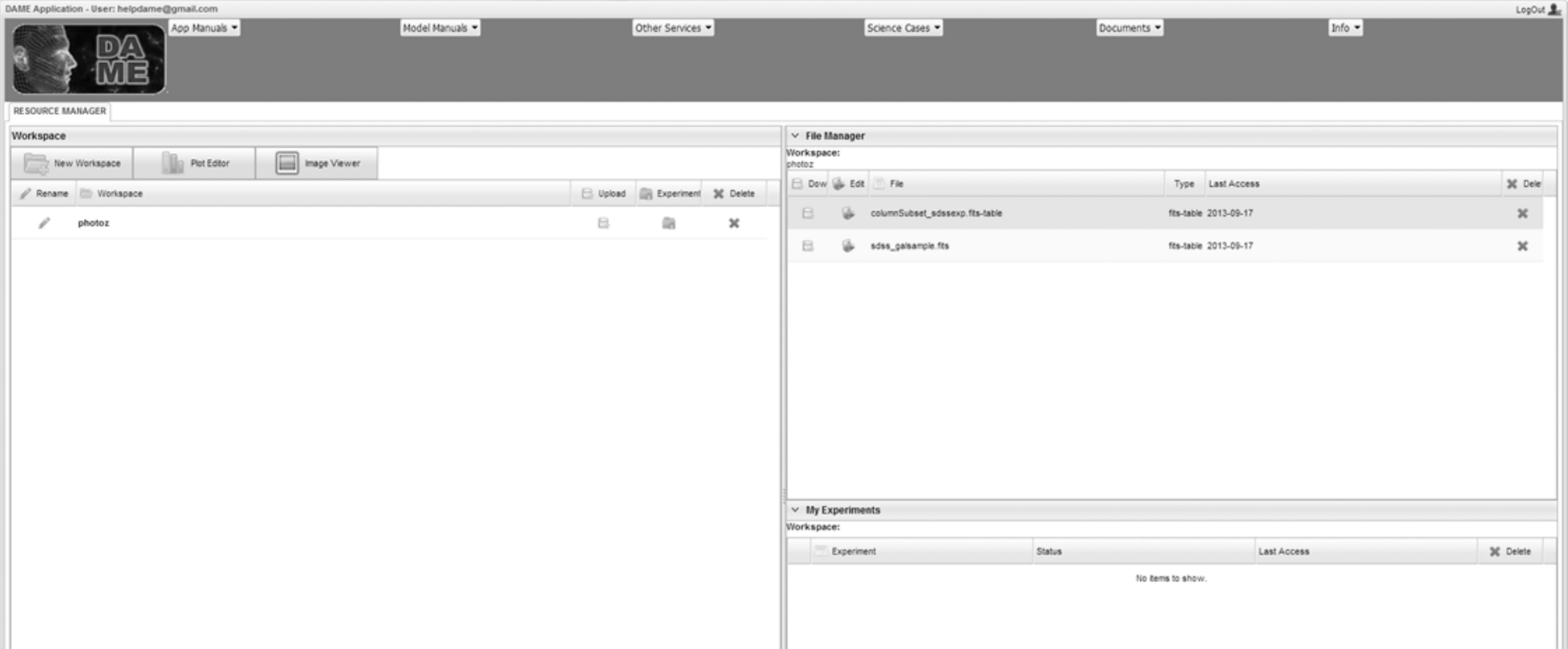}\\
\caption{The \textit{Resource Manager} panel of the DAMEWARE Graphical User Interface, with the three main areas, \textit{Workspace} (left side), \textit{File Manager} (right up) and \textit{My Experiments} (right down), showing the created workspaces, the uploaded data files and the performed experiments, respectively.}\label{fig:gui}
\end{figure*}
\begin{figure*}
\centering
\includegraphics[width=10cm]{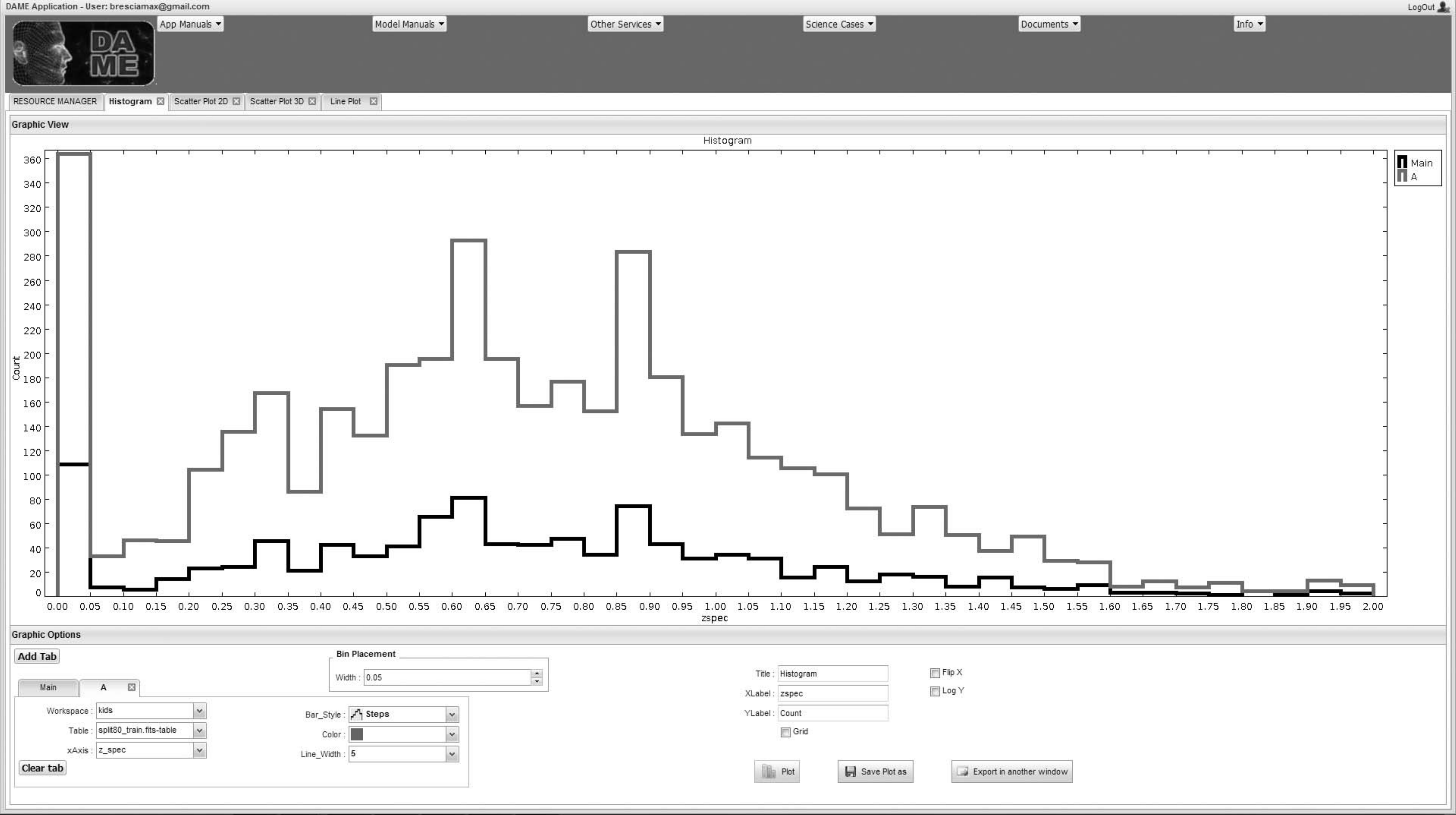}\\(a)

\includegraphics[width=10cm]{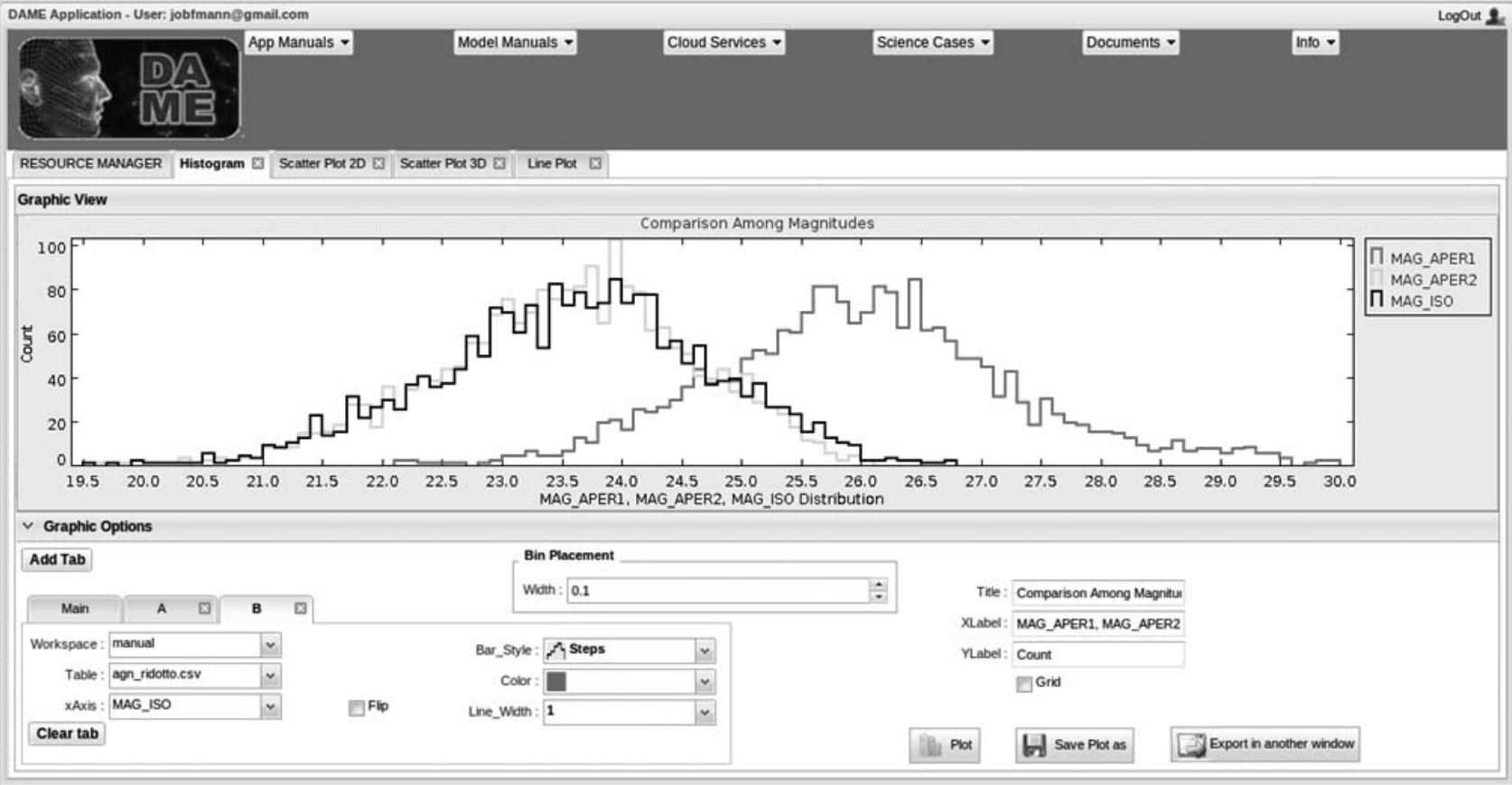}\\(b)

\includegraphics[width=10cm]{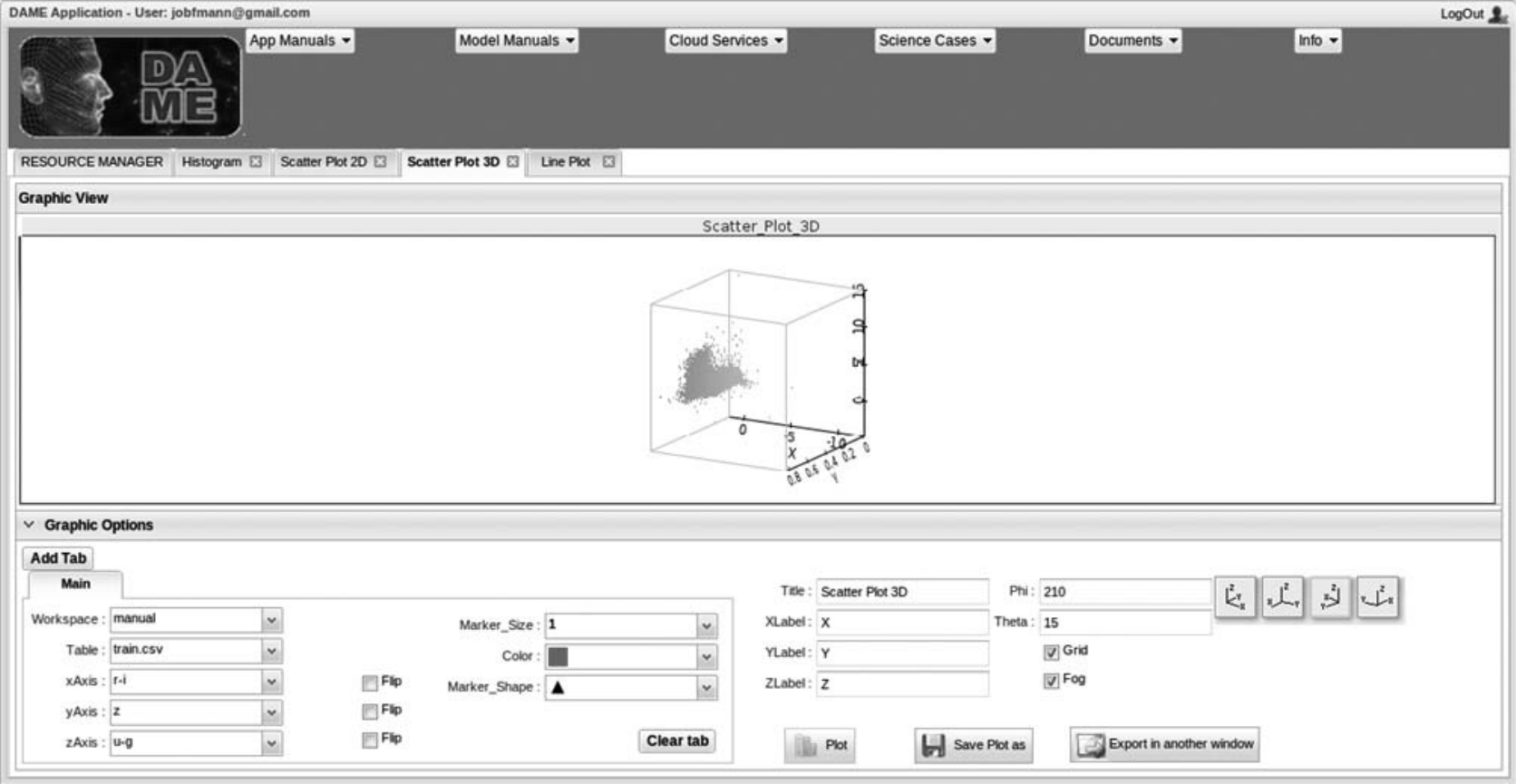}\\(c)

\caption{Panel (a): histogram sample produced by the web application, showing the spectroscopic redshift distribution produced after the splitting of a dataset in two subsets with the \textit{Split by Rows} function.
Panel (b): multi histogram sample showing the photometric distribution of several magnitudes for a given galaxy sample.
Panel (c): 3D scatter plot sample showing the distribution of redshift vs two photometric colors from several magnitudes.}\label{fig:histo}
\end{figure*}

One last consideration: in the case of massive data sets, the user can approach the investigation by extracting
randomly a subset (better several) of data and look at them carefully in order to operate cleaning and to make decisions
about patterns, features and attributes to be organized for future experiments.
The presence of domain experts of course simplifies and reduces this time-consuming preliminary activity.
In any case, the casual user also needs to
be aware that in any successful DM experiment a significant (large) effort must be
put into the pre and post data processing. The literature shows that at least $60\%$ of the time required by a DM application
goes for the data preparation and result verification \citep{cabena1998}.

From what has been said above it should have become clear that data mining experiments require a delicate pre-processing phase
aimed at: i) standardizing the input features; ii) minimizing the effects of missing data; iii)  reducing the dimensionality of
the MPS to a minimum set of independent axes (Reduced Parameter Space or RPS).

\subsection{Machine learning paradigms and models}
There are two standard machine learning paradigms \citep{duda2001}: \textit{supervised} and \textit{unsupervised}.
In the \textit{supervised} paradigm, the dataset needs to contain a sub set of data points (or observations)
for which the user already knows the desired output expressed in terms of categorical classes, numerical
or logical variables, or as a generic observed description of any real problem.
The objects with known output form the so called knowledge base (KB), and provide some level of supervision
since they are used by the learning model to adjust parameters or to make decisions in order to predict the
correct output for new data.
In other words, supervised tasks are the DM equivalent of the usual classification tasks, where the user is required
to divide a sample of objects (for instance galaxies) into classes according to some predefined scheme (e.g. spirals,
ellipticals, lenticulars, etc) learned on the basis of templates or examples.
This is why supervised algorithms are also called \textit{classifiers}.
The outcome is usually a class or category of the examples. Its representation depends on the available KB and on
its intrinsic nature, but in most cases it is based on a series of numerical attributes, organized and submitted in
an homogeneous way.
The success of the learning is usually evaluated by trying out the acquired feature description
on an independent set of data (also named test set), having known output but never submitted to the model before.
Some classifiers are also capable of providing results in a more probabilistic sense, i.e. the probability for a data
point to belong to a given class.
Finally, a classifier can also be used to predict continuous values, a model behavior which is usually called
\emph{regression} \citep{duda2001}.

\begin{figure*}
\centering
\includegraphics[width=10.cm]{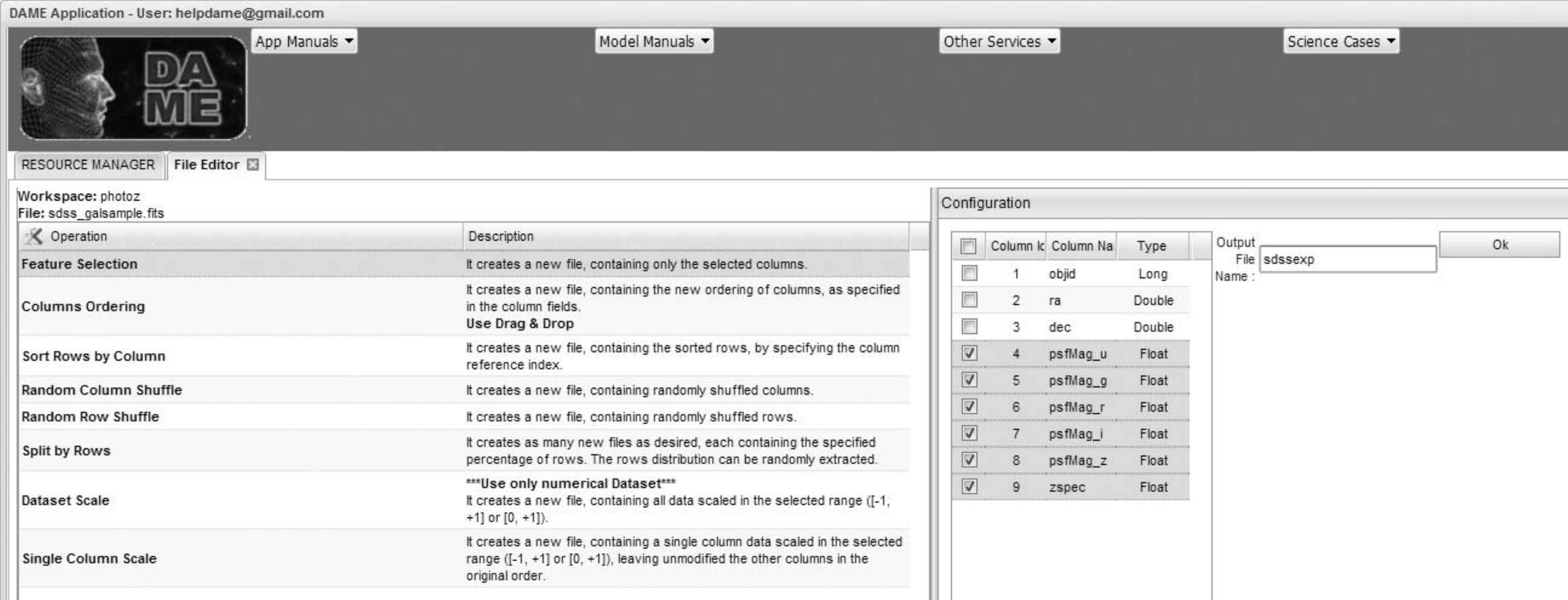}\\(a)

\includegraphics[width=10.cm]{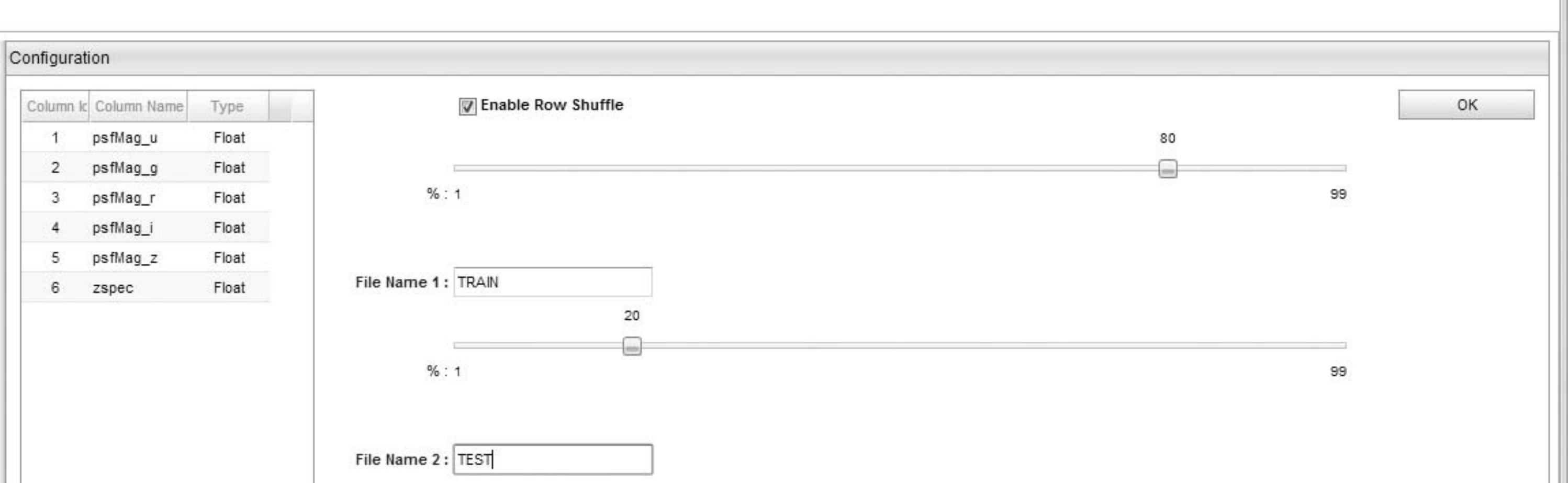}\\ (b)

\caption{
Panel (a): The {\it Feature Selection} option panel, used during the pre-processing phase to extract specific columns from any data table file.
Panel (b): the \emph{Split by Rows} option panel, used during the pre-processing phase to split any data table file into two subsets.}\label{fig:featureextraction}
\end{figure*}
\textit{Unsupervised} algorithms, instead of trying to predict the membership of a datum to one or another a priori defined class,
try to partition the input RPS into disjoint connected regions sharing the same statistical properties.
Each connected region of the partitioned RPS defines what we call a cluster of data points.
In other words, unsupervised algorithms do not learn from examples, but try to create groups or
subsets of the data in which points belonging to a cluster are as similar to each other as much as possible,
by making as large  as possible the difference between different  clusters \citep{haykin1999}.
The success of a clustering process can then be evaluated in terms of human experience, or a posteriori by means of
a second step of the experiment, in which a classification learning process is applied in order to learn an intelligent
mechanism on how new data samples should be clustered.

This basic distinction between \textit{supervised} and \textit{unsupervised} tasks  is reflected in DAMEWARE by the
fact that the choice of any machine learning model is always preceded by selecting its functionality domain.
In other words, since several machine learning models can be used in more than one functionality domain, its choice
is defined by the context in which the exploration of the data is performed.
In what follows we shall adopt the following terminology:

\begin{itemize}
\item \emph{Functionality}: any of the already introduced functional domains, in which the user wants to explore the available
data (such as feature extraction, regression, classification and clustering).
The choice of the functionality target can limit the number of selectable data mining models.
\item \emph{Data mining model}: any of the data mining models integrated in the DAMEWARE framework.
It can be either a supervised machine learning algorithm or an unsupervised one, depending on the available
Knowledge Base (KB), i.e. the set of training or template cases available, and on the scientific target of the experiment.

\item \emph{Experiment}: a complete scientific workflow (including optional pre-processing or preparation of data and post-processing),
starting from the choice of a combination of a data mining model and  the functionality.
\item \emph{Use Case}: for each data mining model, different running cases are exposed to the user.
These can be executed singularly or in a prefixed sequence. Being the models derived from the machine learning paradigm, each one may require a sequence of training (including validation), test and run use cases, in order to perform, respectively, learning, verification and execution phases of the experiment.
\end{itemize}

\begin{table*}
\centering
\footnotesize
\begin{tabular}{|c|l|c|c|}
\hline
\textbf{Model}	& \textbf{Name}                                     & \textbf{Category}	&\textbf{Functionality}\\   \hline
MLPBP 	        & Multi Layer Perceptron with Back Propagation      & Supervised   & Classification, regression\\
FMLPGA          & Fast MLP trained by Genetic Algorithm             & Supervised   & Classification, regression\\
MLPQNA          & MLP with Quasi Newton Approximation               & Supervised   & Classification, regression\\
MLPLEMON        & MLP with Levenberg-Marquardt                      & Supervised   & Classification, regression\\
                & Optimization Network                              &&\\
SVM	            & Support Vector Machine                            & Supervised   & Classification, regression\\
ESOM            & Evolving Self Organizing Maps                     & Unsupervised & Clustering\\
K-Means	        &                                                   & Unsupervised & Clustering\\
SOFM            & Self Organizing Feature Maps                      & Unsupervised &	Clustering\\
SOM             & Self Organizing Maps                              & Unsupervised & Clustering\\
PPS	            & Probabilistic Principal Surfaces                  & Unsupervised &	Feature Extraction\\
\hline
\end{tabular}
\caption{Data mining models and functionalities available in the DAMEWARE framework. Column 1: acronym; column 2: extended name;
column 3: category; column 4: functionality.}
\label{DAME:tab1}
\end{table*}

The functionalities and models currently available in DAMEWARE are listed in Table \ref{DAME:tab1}.
All models are based on the machine learning paradigms and can be grouped into neural networks,
genetic algorithms and other types of self-adaptive methods.
In the category of neural networks, specialized for regression and classification, we list several types
of Multi Layer Perceptrons (MLP, \citealt{mcculloch1943}) with different  learning rules: \textit{(i)} Back Propagation
(MLPBP, \citealt{duda2001}); \textit{(ii)} Genetic Algorithm (FMLPGA), an hybrid model, including genetic programming rules
\citep{mitchell1998}, implemented on both CPU and GPU platforms; \textit{(iii)} Levenberg-Marquardt Optimization Network
(MLPLEMON, \citealt{levenberg1944}) and \textit{(iv)} Quasi Newton Algorithm (MLPQNA, \citealt{shanno1990}).

The neural networks for clustering are the Self Organizing Maps (SOM, \citealt{kohonen2007}), Evolving SOM (ESOM, \citealt{deng2003})
and Self Organizing Feature Maps (SOFM, \citealt{kohonen2007}). These methods are capable to deal also with images (for instance for pixel based clustering)
in the most commonly used formats.

To the category of generic self-adaptive methods belong the Support Vector Machine (SVM, \citealt{chang2011}) for regression and classification,
the K-Means (\citealt{hartigan1979}) for clustering and the Principal Probabilistic Surfaces (PPS, \citealt{chang2001}) for feature extraction.

Depending on the specific experiment and on the execution environment, the use of any model can take place with a more or less advanced
degree of parallelization.
This requirement arises from the fact that all models require the fine tuning of some parameters that cannot be defined a priori, not even by
an experienced user, thus causing the necessity of iterated experiments aimed at finding the best combination.

However, not all the models could be developed under the MPI (Message Passing Interface) paradigm \citep{chu2007} but some models, such as
the FMLPGA, were the F stands for Fast MLPGA, and a general purpose genetic algorithm were also implemented on GPUs in order to speed
up performance \citep{cavuoti2013} moving from multi-core to many-core architectures.

For what scalability is concerned, there are however two different issues: on the one hand the fact that most existing ML methods scale
badly (cf. \cite{gaber2005}) with both increasing number of records and/or of dimensions (i.e., input variables or features); on the other,
the fact that datasets in the multi terabytes range are difficult if not plainly impossible to transfer across the network from the hosting
data centers to the local processing facilities.
During the development period (2008-2011) it become clear that in order to deal with datasets in the Tera and multi-Terabyte range
some changes to the original design had to be introduced even though the true bottle neck was, and still is, in the
fact that very large data sets cannot be transferred over the network and that, in these cases, the whole web app had to be mirrored
in the data centers.
A crucial step in this direction was the implementation of the already mentioned DMPlugin which allows a generic user to configure the I/O interfaces
between his own algorithm and the available infrastructure by generating a wrapper Java which integrates the new model in the suite without
the need to know its internal mechanisms.

\begin{figure*}
\centering
\includegraphics[width=4.cm]{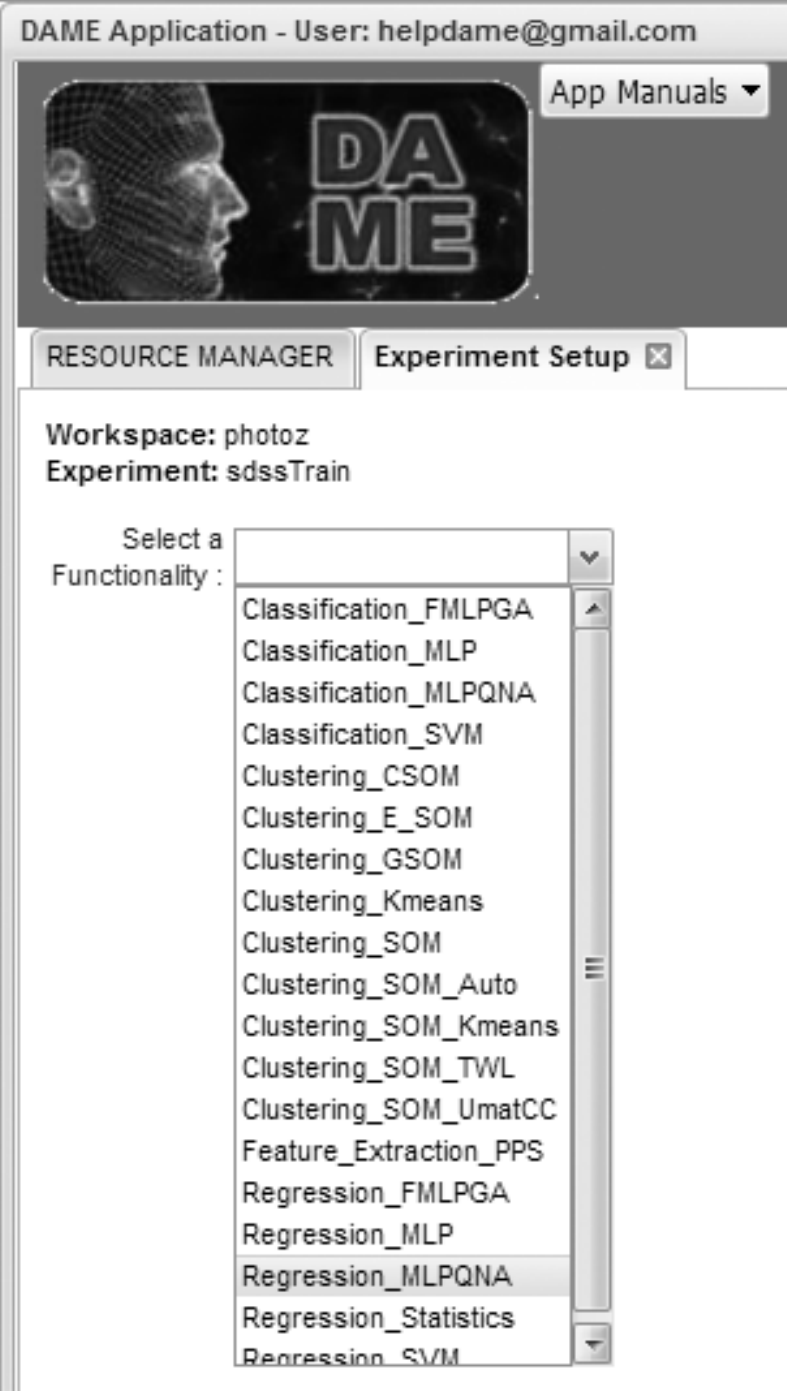}(a)
\includegraphics[width=10.cm]{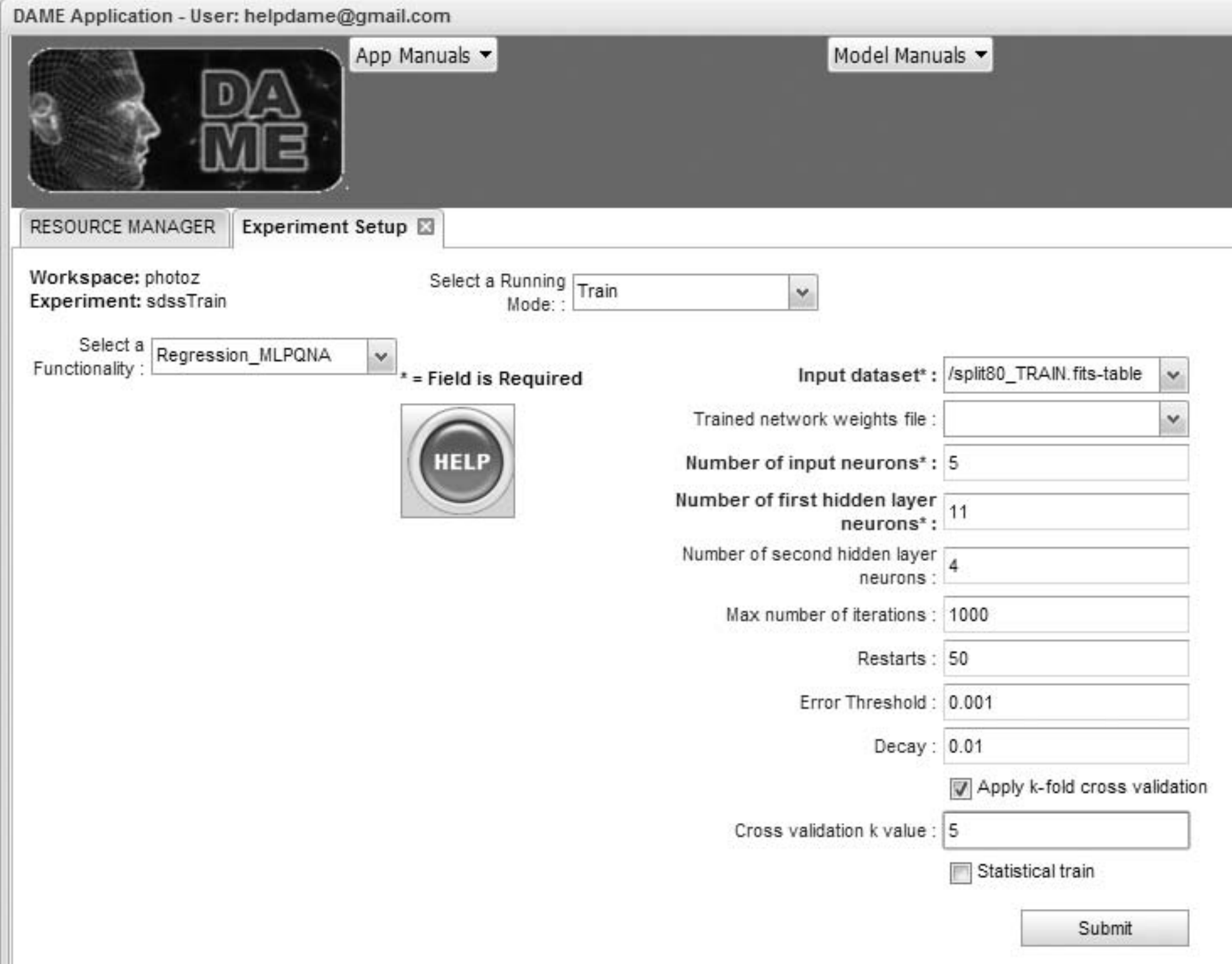}(b)\\
\includegraphics[width=10.cm]{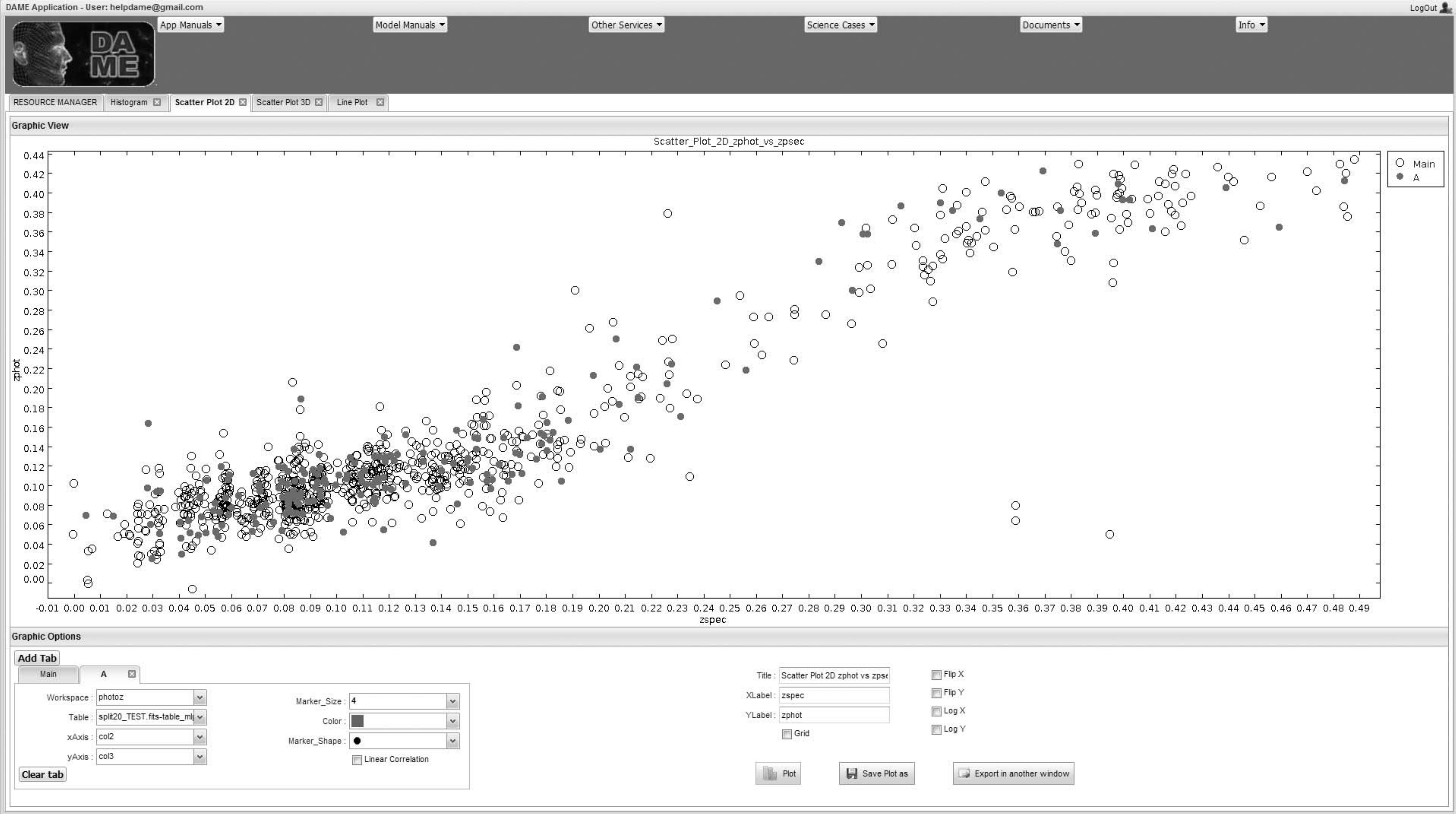}\\(c)

\caption{Panel (a): the selection of the desired couple Functionality-Model for the current experiment.
Panel (b): the train use case configuration panel for the MLPQNA selected model.
Panel (c): the {Scatter Plot 2D} panel, used during the post-processing phase to visualize the scatter plot 2D of zspec vs photo-z. Empty circles are training objects while filled dots are test objects.}\label{fig:experiment}
\end{figure*}

\section{Using DAMEWARE}\label{sect:usingdameware}
In this section we briefly outline the pre-processing and post-processing facilities included in the DAMEWARE platform.

\subsection{Submitting and preparing data}\label{sect:data}

DAMEWARE input data can be in any of the following supported formats: FITS (tabular/image), ASCII (ordinary text, i.e. space separated values),
VOTable (VO compliant XML documents), CSV (Comma Separated Values), JPG, GIF, PNG (image).

In the Graphic User Interface (GUI) the input data belong to a workspace created by the user at the start of any experiment setup process.
All  data are listed in the \textit{File Manager} subwindow (visible in Fig.~\ref{fig:gui}).

Dealing with machine learning methods, starting from an original data file, a typical pre-processing phase
consists of the preparation of several different files to be submitted as input for training, testing and validation of the chosen algorithm. The pre-processing features available in DAMEWARE are (cf. also Fig.~\ref{fig:featureextraction}a):

\begin{itemize}
\item	\textit{Feature Selection}: it allows the creation of a new file containing only user selected columns from the original input file;
\item	\textit{Columns Ordering}: it creates a new file containing the user specified order of columns;
\item	\textit{Sort Rows by Column}: it allows the creation of a new file containing rows sorted on a user selected row index;
\item	\textit{Column Shuffle}: it creates a new file containing shuffled columns;
\item	\textit{Row Shuffle}: it creates a new file containing shuffled rows;
\item	\textit{Split by Rows}: it allows the creation of two new files containing the user specified percentage of rows (the row distribution can be randomly extracted);
\item	\textit{Dataset Scale}: it creates a new file with normalized values in a chosen range ($[-1, +1]$ or $[0, 1]$);
\item	\textit{Single Column scale}: it allows the creation of a new file with the values of a selected normalized column in a chosen range, leaving the rest of columns unchanged;
\end{itemize}

Other, less DM oriented, tasks need to be execute outside of DAMEWARE, using programs such as TOPCAT \citep{taylor2005}.

\subsection{Inspecting results}\label{sect:visualization}

Any outcome of a machine learning based experiment, originates from an inductive process, hence needs to be post-processed.
Post-processing always helps to investigate, as well as to fine tune, the acquired knowledge.
It usually requires both statistics and graphical representations.
DAMEWARE provides both kinds of tools.

For instance, classification confusion matrices \citep{brescia2012b} and regression residual analysis are the available statistical tools for supervised models.
A series of graphical plots enables instead the investigation of the outcome of unsupervised (clustering and feature extraction) experiments as well as
the inspection of particular trends of a generic data table or  the viewing of a generic astronomical image.

\noindent The graphical options selectable by the user are:
\begin{itemize}
\item multi-column histograms;
\item multi-tab scatter plot 2D;
\item multi-tab scatter plot 3D;
\item multi-column line plot;
\item visualization of the most common image types (gif, jpeg, fits, png).
\end{itemize}
\begin{figure*}
\centering
    \includegraphics[width=12.cm]{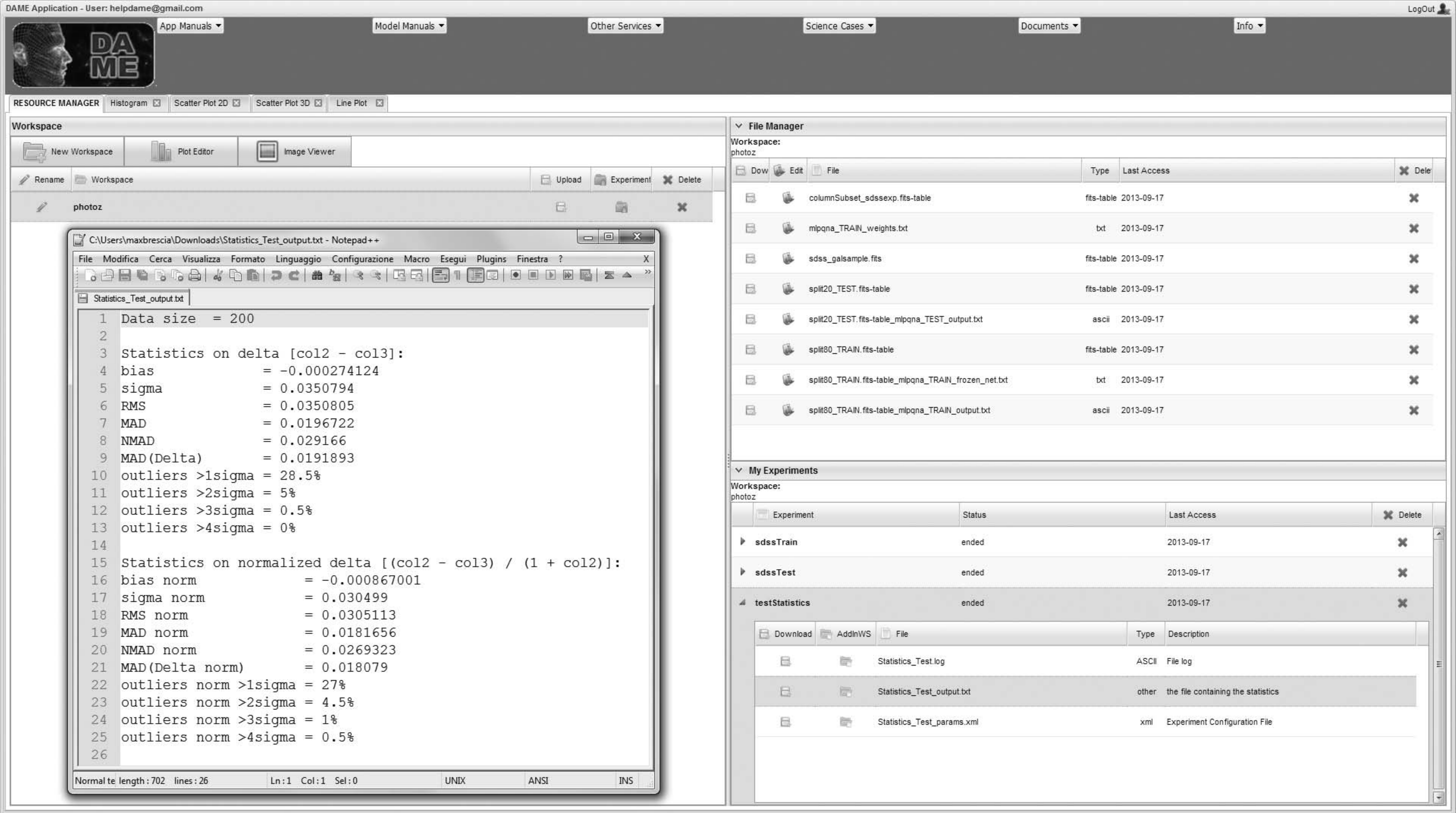}
\caption{The output of the test experiment with also the residual statistics report. On the right side the \textit{File Manager}
and the \textit{My Experiments} areas are shown.}\label{fig:statistics}
\end{figure*}

Some examples of graphical output are shown, respectively, in Fig.~\ref{fig:histo} a,b,c and
Fig.~\ref{fig:lineplot}.

\section{Scientific applications of DAMEWARE}\label{sect:science}
During the development phase, DAMEWARE has been tested on many science cases which have led to significant results,
published in several scientific papers.
In order to better exemplify the potential application of DAMEWARE, in what follows we shall briefly outline some
recent applications and results.

\subsection{Classification tasks}
A typical astronomical problem tackled in the past with automatic tools is the so called star/galaxy classification task
which, at least until the late 90's should have been more correctly described as disentangling unresolved (i.e. point like)
and spatially resolved (i.e. extended) objects.
Nowadays, even though the basic problem is always the same, the possibility to use multi-band information adds an additional
level of complexity and allows one to disentangle not only resolved vs unresolved objects, but also unresolved
extragalactic objects against unresolved galactic ones.
We performed a particular S/G classification experiment using DAMEWARE to identify candidate globular clusters in the
halo of the galaxy NGC1399, disentangling them from background galaxies and foreground galactic stars.
In this classification task, the KB consisted of a set of bona fide globular clusters selected in a small portion of the field
on the basis of color-color diagrams.
This selection criterion, while very effective, requires multi-band observations and high angular resolution.
The aim of our DM experiment was to use the KB to train a model to classify candidate globular clusters using single band
observations obtained with the Hubble Space Telescope (thus covering a much larger field of view with respect to the multi-band
data).
DAMEWARE allowed us to test and compare different DM models and to choose the optimal one.
In particular, three different versions of MLPs, genetic algorithms and SVM were used on the same data set \citep{brescia2012}.
The best results were obtained with the MLPQNA leading to a class accuracy of 98.3\%, a completeness of 97.8\% and a contamination of 1.8\%.

Another classification problem tackled with DAMEWARE was to disentangle on photometric grounds only
galaxies hosting active nuclei (AGN) from normal (i.e. non active) ones and to try to separate AGNs into broad
phenomenological types such as Seyfert I, Seyfert II and LINERs \citep{cavuoti2013b}.
In this case, the KB was more complex since it was assembled from different catalogues (all built on the basis of spectroscopic information).
Also in this case DAMEWARE was used to test different classification models (different implementations of MLP and SVM).
More specifically, we addressed three different classification problems:
i) the separation of AGNs from non-AGNs,
ii) Seyfert I from Seyfert II, and
iii) Seyfert from LINERs.
In terms of classification efficiency, the results indicated that our methods performed fairly well ($\sim76.5\%$) when applied
to the problem of the classification of AGNs vs non-AGNs, while the performances in the finer classification of Seyfert vs LINERs
resulted $\sim78\%$ and $\sim81\%$ in the case Seyfert I vs Seyfert II.
The relatively low percentages of succesfull classification are compatible with what is usually achieved in the literature
and reflect the ambiguities present in the KB. The resulting catalogue, containing more than 3.2 million candidate AGNs is available
on-line on the VizieR service \citep{catagn}.

\subsection{Regression for photometric redshifts}

The evaluation of photometric redshifts (hereinafter photo-z) is among the first and most common problems dealt by astronomers
using machine learning or data mining methods.
Photometric redshifts offer a viable and less demanding in terms of precious observing time, alternative to spectroscopy based
techniques, to derive the redshifts of large samples of galaxies.
In practice, the problem consists of finding the unknown function which maps a photometric set of features (magnitudes and/or colors)
into the redshift space and many different techniques and approaches have been developed \citep{hildebrandt2010}.
When a consistent fraction of the objects with spectroscopic redshifts exists, the problem can be approached as a DM regression
problem, where the a priori knowledge (i.e. the spectroscopic redshifts forming the KB) is used to uncover the mapping function.
This function can then be used to derive photo-z for objects which have no spectroscopic information.

Without entering into much detail, which can be found in the literature quoted below and in the references therein, we just
summarize a few salient aspects tested in many experiments done on different KBs, often composed through accurate cross-matching
among public surveys, such as SDSS for galaxies \citep{brescia2014}, UKIDSS, SDSS, GALEX and WISE for quasars \citep{brescia2013},
GOODS-North for the PHAT1 contest \citep{hildebrandt2010,cavuoti2012} and CLASH-VLT data for galaxies \citep{biviano2013}.
Other photo-z prediction experiments are in progress as preparatory work for the Euclid Mission
\citep{laureijs2011} and the KIDS\footnote{\url{http://www.astro-wise.org/projects/KIDS/}} survey projects.

While referring the interested reader to the above quoted papers and to Sect. \ref{sect:app} for details, we just notice that in all
these experiments we exploited a DM functionality which appears to be relevant for a better understanding of the feature selection
possibilities offered by DAMEWARE.
In \citealt{brescia2013} it is exemplified how the use of feature selection, outlined in Sect. \ref{sect:DM}, could be used to reduce
the number of significant input parameters from the initial $43$ to only $15$, with no loss in regression accuracy and with a huge
improvement in computing time.

\section{A template science case}\label{sect:app}

In this section, which must be regarded as a sort of a tutorial, we show how to use the MLPQNA model in DAMEWARE to evaluate photometric
redshifts for a sample of objects which was available for the PHAT$1$ contest \citep{hildebrandt2010, cavuoti2012}.

According to \cite{hildebrandt2010}, due to the extremely small KB of spectroscopic redshifts, PHAT1 provides a quite complex
environment where to test photo-z
methods.
The PHAT$1$ dataset consists of photometric observations, both from ground and space instruments (GOODS-North; \citealt{giavalisco2004}),
complemented by additional data in other bands derived from \cite{capak2004}. The final dataset covers the full UV-IR range and includes $18$
bands: U (from KPNO), B, V, R, I, Z (from SUBARU), F435W, F606W, F775W, F850LP (from HST-ACS), J, H (from ULBCAM), HK (from QUIRC), K (from WIRC),
and $3.6$, $4.5$, $5.8$, and $8.0$ $\mu$ (from IRAC Spitzer).
The photometric dataset was then cross correlated with spectroscopic data from \cite{cowie2004}, \cite{wirth2004}, \cite{treu2005}, and \cite{reddy2006}.
Therefore, the final PHAT$1$ dataset consists of $1984$ objects with $18$-band photometry and more than one quarter of the objects ($515$), with
accurate spectroscopic redshifts. Details about the machine learning model MLPQNA used in this context, can be found in \cite{cavuoti2012}.

In the following we will just describe more practical aspects of the workflow along the experiment development.
Details and available data of the experiments can be found on the DAME web site\footnote{\url{http://dame.dsf.unina.it/dame_photoz.html\#phat}}.

All the following experiment phases are intended to be performed after having successfully completed the access and preliminary steps on the DAMEWARE web application:
\begin{enumerate}
\item Login into the web application using your username and the password obtained after the registration procedure;
\item Create a new workspace using the specific button in the main GUI window (see Fig.~\ref{fig:gui});
\item Upload the original data files containing the complete KB (it can be loaded from user local machine or from remote web address
by providing the URL);
\end{enumerate}

Let us start from the construction of the knowledge base needed for training, validation and test. For supervised methods it is
common practice to split the KB into at least three disjoint subsets: one (training set) to be used for training purposes,
i.e. to teach the method how to perform the regression; the second one (validation set) to check against the possible loss of
generalization capabilities (also known as overfitting); and the third one (test set) needed to evaluate the performances of the
model.

As a rule of thumb, these sets should be populated with $60\%$, $20\%$ and $20\%$ of the
objects in the KB. In order to ensure a proper coverage of the MPS, objects in the KB are divided up among the three datasets by random extraction, and usually this process is iterated several times in order to minimize the biases introduced by fluctuations in the coverage of the PS.
In the case of MLPQNA described here, we used the \textit{leave-one-out} k-fold cross-validation (cf. \citealt{geisser1975}) to minimize the size of the validation set.
Training and validation were therefore performed together using $\sim80\%$ of the objects as a training set and the remaining $\sim 20\%$ as test set
(in practice $400$ records in the training set and $115$ in the test set).

To ensure proper coverage of the MPS, we checked that the randomly extracted populations had a spec-z distribution compatible with that of the whole KB.
The automated process of cross-validation ($K=10$) was then done by performing ten different training runs with the following procedure:
\textit{(i)} the training set was split into ten random subsets, each one containing  $10\%$ of the objects;
\textit{(ii)} at each run we used 9 of the small datasets for the training and the remaining one for validation.

The second phase of the experiment consists of the analysis of data features and missing data.
As already mentioned, the presence of features with a large fraction of NaNs can seriously affect the performance of
a given model and lower the accuracy or the generalization capabilities of a specific model.
It is therefore good practice to analyze the performance of a specific model in presence of features with large fractions of NaNs.
This procedure is strictly related to the \textit{feature selection} phase which consists in evaluating the significance of individual
features to the solution of a specific problem.
We wish to recall that, as a rule of thumb, ``feature selection methods belong to two large families: ``filter modeling'' and ``wrapper modeling''.
The first group includes methods based either on specific models (such as Principal Component Analysis or PCA, etc.) or on
statistical filtering which require some a-priori knowledge on the data model, while the second group uses the machine learning method itself
to assess the significance of individual features.
It is also necessary to underline that especially in the presence of small datasets, there is a need for  compromise: on the one hand, it is
necessary to minimize the effects of NaNs; on the other, it is not possible to simply remove each record containing NaNs, because otherwise
too much information would be lost.
Details of the data analysis and feature selection performed on the PHAT1 dataset are described in \citealt{cavuoti2012}.

The construction and manipulation of the data sets for the above phases can be performed in DAMEWARE, after user access, through the data editing
options presented in Sect.~\ref{sect:data}.

Starting from the availability in any user workspace of the data sets (train and test set) containing columns related to the input features
(photometric magnitudes, colors or fluxes of all selected/available bands) and reference output (spectroscopic redshift) for all objects,
the train and test phases can be done by performing the following series of steps:

\begin{enumerate}
\item upload in the created workspace the train, test and run data sets from the website\footnote{\url{http://dame.dsf.unina.it/dame_photoz.html\#phat}}.
Here for \textsl{run} data set we mean the whole data set containing photometric information only to which the network will be applied at the end
of the training+test procedure.
\item create the files to be used for the train and test phases, by using the \textit{Feature Selection} option in the editing menu (enabled by
clicking the \textit{Edit} icon close to each file). In particular the three columns labeled as $ID$, $18-band$ and $14-band$, respectively, should
be removed in both train and test sets, because not involved as input features in the experiment;
\item Create a new train experiment (by clicking the icon named \textit{Experiment} close to the workspace name). Select the desired ML model having
the regression functionality among those available in the experiment type list (in this case Regression\_MLPQNA), (see for example the
Fig.~\ref{fig:experiment}a);
\item Choose the use case \textit{train} among those made available in the selected model list (the train use case is always the first required for
new experiments);
\item Configure all required model and experiment type parameters (you can also follow the suggestions obtained by pressing the \textit{Help button},
as shown in Fig.~\ref{fig:experiment}b).
The optimal setup of the parameters is usually found by following a trial and error procedure but, in this case, if the user does not want to run the
numerous experiment needed, he can use the set of parameters defined in \citealt{cavuoti2012};
\item Launch the train experiment and wait until the status \textit{ended} appears in the \textit{My Experiments} panel (after the launch the user can disconnect from the web application);
\item  At the end of the experiment, move the internal configuration files of the trained model (depending on the model used) from the experiment output list to the \textit{File Manager} area of the current workspace (by pressing the \textit{AddInWS} button nearby). For instance, if MLPQNA was selected, the weight file is named as \textit{mlpqna\_TRAIN\_weights.txt} and the network setup file as \textit{[inputfilename]\_mlpqna\_TRAIN\_frozen\_net.txt}, where the prefix depends on the specific input file name;
\item  Create a new test experiment in the same workspace and choose the use case \textit{test}, configure its parameters (a subset of those already used in the
train setup with the obvious difference that in this case the input must be the test set previously prepared in the workspace and by selecting the weight and
network setup files, obtained ath the end of training) and launch it;
\item At the end of the test case experiment, move the output table file in the \textit{File Manager} area;
\item Evaluate the results by performing some post-processing steps. For instance:
\begin{itemize}
\item Visualize the scatter plot (zspec vs photo-z) by pressing the menu button \textit{Plot Editor};
\item Select the sub-tab \textit{Scatter Plot 2D}, load the input file, configure the plot options and create it (an example is shown in Fig.~\ref{fig:experiment}c);
\item Create a new statistics experiment by selecting the Regression-statistics experiment option and launch it by submitting the test output file as input data and
columns $2$ and $3$ as references for the calculations;
\item Download and visualize on your local machine the residual statistics output file;
\end{itemize}
\end{enumerate}

In the case that the user has not adopted the suggested values but is trying to derive the optimal setup on a \textit{trial-and-error} basis, the whole
procedure needs to be repeated many times by varying the setup parameters and by comparing the resulted statistics.

After completing the sequence of \textit{train, test, statistics, plotting} experiments, the user can use the trained model to produce photo-z for the objects
without spectroscopic redshift, i.e. the \textit{run} set. To do this, he needs a third file containing the \textit{photometric only} objects, pruned of all
the data with incomplete information and filtered accordingly to the same photometric selection criteria mentioned above.
Using the trained network, the user must choose the use case \textit{run}, configure its parameters (a subset of those already used in the train and test setup
with the obvious difference that in this case the input must be the \textit{run} set previously prepared in the workspace) and launch it.
The output file will contain the estimated photo-z for all given objects.

\section{Future developments}
As discussed in the previous sections, DAMEWARE is fully capable to deal with most existing data sets but,
due to the limitations imposed by data transfer over the network, it is clear that the future of any DAMEWARE-like service
will depend on the capability of moving the data mining applications
to the data centers hosting the data themselves.

The VO community has already designed web based protocols for application interoperability (such as the Web Samp Connector), which
solve some problems but still require to exchange data between application sites \citep{derrierre2010,goodman2012a}.
From a conceptual point of view, the possible interoperability scenarios are (hereinafter DA stands for Desktop Application and WA for Web Application):

\begin{enumerate}
\item	$DA1 \Leftrightarrow DA2$ (data + application bi-directional flow)
\begin{enumerate}
\item	Full interoperability between DAs;
\item	Local user desktop fully involved (requires computing power);
\end{enumerate}
\item	$DA \Leftrightarrow WA$ (data + application bi-directional flow)

\begin{enumerate}
\item   Full $WA \Rightarrow DA$ interoperability;
\item	Partial $DA \Rightarrow WA$ interoperability (such as remote file storing);
\item	MDS must be moved between local and remote applications;
\item	user desktop partially involved (requires minor computing and storage power);
\end{enumerate}

\item	WA $\Leftrightarrow$ WA (data + application bi-directional flow)
\begin{enumerate}
\item   Except from URI exchange, no interoperability and different accounting policy;
\item	MDS must be moved between remote apps (but larger bandwidth);
\item	No local computing power required;
\end{enumerate}
\end{enumerate}

\noindent All these mechanisms are however, just partial solutions since they still require to exchange data over the web between application sites.

Building upon the DAMEWARE experience and in particular upon the experience gained with the implementation of the DMPlugin resource,
we decided to investigate a new concept called \textit{Hydra}\footnote{The name was inspired by the Greek mythological monster called
\textit{Lernaean Hydra}, the ancient nameless serpent-like water beast, with reptilian traits that possessed many independent but equally
functional heads}.
Within this concept, we started by designing a prototype of a standardized web application repository, named \textit{HEAD}.
A HEAD (Hosting Environment Application Dock) cloud is in practice a group of standardized software containers of data mining
models and tools, to be installed and deployed in a pre-existing data warehouse.
In such scenario, the various HEADs can be different in terms of number and type of models made available at the time of installation.
This is essentially because any hosting data center could require specific kinds of data mining and analysis tools, strictly related
with their specific type of data and specific knowledge search types.
All HEADs, however, would be based on a pre-designed set of
standards, completely describing their interaction with external environment, application plugin and execution procedures and therefore
would be identical in terms of internal structure and I/O interfaces.
If two generic data warehouses host two different HEADs on their site, they are able to engage the mining application interoperability
mechanism by exchanging algorithms and tool packages on demand.

On a specific request, the mechanism will engage an automatic procedure which moves applications, organized under the
form of small packages (a few MB in the worst case), through the Web from a HEAD source to a HEAD destination, installs them and
makes the receiving HEAD able to execute the imported model on local data.

Of course, such strategy requires a well-defined design approach, in order to provide a suitable set of standards and common rules
to build and codify the internal structure of HEADs and data mining applications, such as for example any kind of rules like PMML,
Predictive Model Markup Language \citep{guazzelli2009}.
These standards can be in principle designed to maintain and preserve the compliance with data representation rules and protocols
already defined and currently operative in a particular scientific community (such as VO in Astronomy).

In order to fine tune the Hydra concepts, we recently approached the design and development of a desktop
application prototype capable to deal with general classification and regression problems, but fine tuned to
tackle specific astrophysical problems.
The first example being the PhotoRApToR\footnote{\url{http://dame.dsf.unina.it/dame_photoz.html\#photoraptor}}
(PHOTOmetric Research APplication To Redshifts) App, freely downloadable at the DAMEWARE project website,
for the supervised prediction of photometric redshifts \citep{destefano2014}.

\section{Conclusions}\label{sect:conclusion}
The DAMEWARE project started in $2007$ and was released in $2013$.
These have been momentous years for astronomy which has become a data rich science and is now coping
with data problems whose full extension could just be imagined a decade ago.

We therefore think it useful to briefly account for some lessons learned in the making of a project which to
our knowledge is among the very few projects aimed at providing the astronomical community with a user friendly tool capable
to perform extensive data mining experiments on massive data sets.
Some of the considerations below arise from the experience gathered in answering frequent questions raised by the community of users and
might prove of general interest.

An important aspect, revealed through the DAMEWARE experience, regards the computing latency time required by large data mining experiments.
For instance, the completion of the set of experiments described in \cite{brescia2013} required several weeks of computing time on a
multicore processor.

The harder problem for the future will be heterogeneity of platforms, data and applications, rather than simply the scale of the deployed resources.
The goal should be to allow scientists to explore the data easily, with sufficient processing power for any desired algorithm to efficiently process it.
Most existing ML methods scale badly with both increasing number of records and/or of dimensionality (i.e., input variables or features).
In other words, the very richness of astronomical data sets makes them difficult to analyze.
This can be circumvented by extracting subsets of data, performing the training and validation of the methods on these more manageable data subsets,
and then extrapolating the results to the whole data set.
This approach obviously does not use the full informational content of the data sets, and may introduce biases which are often difficult to control.
Typically, a lengthy fine tuning procedure is needed for such sub-sampling experiments, which may require tens or sometimes hundreds of experiments
to be performed in order to identify and optimize, for the problem in hand, the DM method or its architecture and parameter setup.

The DAMEWARE resource was designed by taking all these issues into account, thus including the parallel computing facilities, based on GPGPU hardware
and CUDA+OpenACC software paradigms \citep{nvidia2012}, applied to most expensive models, like hybrid architectures (neural networks with genetic algorithms).
The speedup gain obtained by executing such models on the parallel platforms ensures the scalability of our algorithms and makes feasible the data mining
process with them on huge data sets. So far, all future astrophysical data mining resources require a massive exploitation of such paradigms.

With the release of the current version, DAMEWARE has concluded the test phase and has become fully operational. However, besides debugging and the addition
of further data mining models and methods, no major revisions and/or additions are foreseen.

\section*{Acknowledgments}

\noindent The authors wish to thank the many students who collaborated one way or the other to the project.
Many of them played such an important role to earn a place in the authors list some others made small but significant contributions.
To all of them we are very grateful. The authors wish also to thank the anonymous referee for his/her many helpful suggestions.\\
\noindent DAMEWARE has been a multiyear project funded by many sponsors. The Italian Ministry of Foreign Affairs through a bilateral Great Relevance
Italy-USA project; the European funded VO-Tech (Virtual Observatory Technological Infrastructure) 6-th European FW project; the Department of Physics
and the Polo della Scienza e Della Tecnologia of the University Federico II in Napoli.
GL, MB and SC acknowledge financial support by the Project F.A.R.O., $3^{rd}$ call by the University Federico II of Naples.
GL wish also to thank the financial support by the PRIN-MIUR 2011, \textit{Cosmology with the Euclid space mission}.
SGD, CD, AAM, and MJG acknowledge a partial support from the NSF grants AST-0834235 and IIS-1118041, and the NASA grant
08-AISR08-0085.\\
\noindent Some of the work reported here benefited from the discussions which took place during a study and the workshops organized by the Keck Institute for
Space Studies at Caltech.

\end{document}